\documentstyle[preprint,aps]{revtex}

\begin{document}

%\draft         % PRINTS PACS NUMBER
\preprint{MADPH-95-893 IFUSP-P 1160 }

\title{Anomalous {\boldmath$g^Z_5$} Coupling at {\boldmath$\gamma\gamma$}
Colliders}

\author{O.\ J.\ P.\ \'Eboli \cite{ojpe},}

\address{Instituto de F\'{\i}sica,
Universidade de S\~ao Paulo, \\
C.\ P.\  66318,  05389-970 S\~ao Paulo, Brazil}

\author{M.\ B.\ Magro \cite{mbm}, and }

\address{Departament  de F\'{\i}sica Te\`orica, Universitat de Val\`encia
and IFIC, \\
46100 Burjassot, Val\`encia, Spain}

\author{P.\ G.\ Mercadante \cite{pgm} }

\address{Physics Department, University of Wisconsin, \\
Madison, WI 53706, USA}

\date{\today}
\maketitle

\begin{abstract}

  We study the constraints on the anomalous coupling $g^Z_5$ that can
 be obtained from the analysis of the reaction $\gamma\gamma
 \rightarrow W^+ W^- Z$ at future linear $e^+e^-$ colliders. We find
 out that a $0.5$ ($1$) TeV $e^+e^-$ collider operating in the
 $\gamma\gamma$ mode can probe values of $g_5^Z$ of the order of
 $0.15$ ($4.5 \times 10^{-2}$) for an integrated luminosity of $10$
 fb$^{-1}$.  This shows that the ability to search for this anomalous
 interaction of the $\gamma\gamma$ mode is better than the one of the
 usual $e^+e^-$ mode, and it is similar to the ability of the
 $e\gamma$ mode.

\end{abstract}

\vskip .2in

\begin{center}
{\em To be published in Physical Review D1}
\end{center}

%\pacs{XXX}

\newpage

%**********
% PAPER
%**********
\section{Introduction}
\label{sec:int}

The mechanism for breaking the symmetry of the electroweak interactions
has not been directly accessed in experiments thus far. One possibility
is that the Higgs boson is so heavy that it will not be produced even in
the next generation of colliders. In this case and in other strongly
interacting symmetry-breaking scenarios, it is interesting to
parametrize the symmetry-breaking sector in a model independent way
through the use of chiral lagrangians \cite{appel}. In this approach,
the low energy effects of new physics are represented by  an infinite
tower of non-renormalizable  effective operators which are consistent
with the $SU(2)_L \otimes U(1)_Y$ symmetry of the standard model (SM).

The lowest order chiral Lagrangian exhibits an universal behavior  for
the dynamics of the electroweak interactions, being independent of the
details of the mechanism of symmetry-breaking. However, at the
next-to-leading order in the chiral expansion, there are 14 effective
operators whose coefficients are dictated by the underlying dynamics.
Among the next-to-leading operators, there is only one that is CP
conserving but parity violating \cite{apwu}. This operator also breaks
the custodial $SU(2)_C$ symmetry \cite{cust} and is given by
\begin{equation}
{\cal L}_{11} = \alpha_{11} g~ \epsilon^{\alpha \beta \mu \nu}
\text{Tr} \left ( \tau^3 U^\dagger D_\mu U \right )~
\text{Tr} \left ( U^\dagger W_{\alpha\beta} D_\nu U \right )
\; ,
\label{lag:gz5}
\end{equation}
where the dimensionless unitary unimodular matrix $U = \exp (  i \xi^a
\tau^a / v^2 )$ contains the would-be Goldstone bosons $\xi^a$,  $v
\simeq 246$ GeV is the symmetry-breaking scale, the $SU(2)_L \otimes
U(1)_Y$ covariant derivative is
\begin{equation} D_\mu U = \partial_\mu U + i~ \frac{g}{2}~ W^j_\mu \tau^j U
 - i~ \frac{g^\prime}{2}~ B_\mu U \tau^3
\; ,
\end{equation}
and the field strength tensors are written in terms of $W_\mu = W^j_\mu
\tau^j$
\begin{eqnarray}
W_{\mu\nu} &=& \frac{1}{2} \left ( \partial_\mu W_\nu
- \partial_\nu W_\mu + \frac{i}{2} [ W_\mu, W_\nu] \right )
\; , \\
B_{\mu\nu} &=& \frac{1}{2} \left ( \partial_\mu  B_\nu
- \partial_\nu B_\mu \right ) \tau^3
\; .
\end{eqnarray}

The physical content of the above operator is more transparent in the
unitary gauge, $U = 1$, where the effective Lagrangian (\ref{lag:gz5})
gives rise to anomalous contributions to the triple vertex $W^+ W^- Z$
and to the four-gauge-boson vertex $W^+ W^- Z \gamma$. In the standard
notation of Ref.\ \cite{dieter}, we have the correspondence for the
triple gauge-boson vertex
\begin{equation}
g^Z_5 =  \frac{e^2}{s_W^2 c_W^2} \alpha_{11} \; ,
\end{equation}
where we denote the sine (cosine) of the weak mixing angle by $s_W$
($c_W$). The expected size of $g^Z_5$ depends upon whether or not the
underlying dynamics respects the custodial symmetry. In models
with a  custodial symmetry, $g^Z_5$ should be of the order of $10^{-4}$,
while for models without this symmetry we expect $g^Z_5 \sim 10^{-2}$.

At low energies, the bounds on this operator come from one-loop
contributions to meson decays and to the vertex $ Z f \bar{f}$. From the study
of the decay $K_L \rightarrow \mu^+ \mu^-$, we obtain limits of the order
 $g^Z_5 \lesssim 1$ \cite{kl}, while the precise measurements of
the $Z$ flavor diagonal couplings imply that $g^Z_5 \lesssim 0.04$
\cite{sdaw}. These bounds are obtained using the naturalness assumption
that no cancellations take place  between contributions from different
anomalous interactions. However, a closer look at the interaction
(\ref{lag:gz5}) reveals that it is momentum dependent, and consequently
it can be better studied directly in processes at high energies.

The Next Linear $e^+e^-$ Collider (NLC) \cite{pal} will reach a
center-of-mass energy between 500 and 2000 GeV with an yearly integrated
luminosity of at least $10$ fb$^{-1}$. An interesting feature of this
new machine is the possibility of transforming an electron beam into a
photon one through the laser backscattering mechanism \cite{las0,laser}.
 This process will allow the NLC to operate in three different modes,
$e^+e^-$, $e\gamma$, and $\gamma\gamma$, opening up the opportunity for
a wider search for new physics. However, it is important to stress that
the collider can operate in only one of its three modes at a given time,
therefore, it is imperative to study comparatively the different
features of each of these setups.

Previously, the phenomenological implications of the operator
(\ref{lag:gz5}) to the reaction $e^+e^- \rightarrow W^+ W^- Z $ at high
energies were analyzed in Ref.\ \cite{sdaw:ee}, which showed that it is
possible to obtain limits of the order of $g^Z_5 \lesssim 0.3$ for a
center-of-mass energy of 500 GeV. However, this sensitivity to $g^Z_5$
can only be achieved for a high degree of $e^-$ polarization. This
interaction was also studied in $e\gamma$ collisions in Ref.\
\cite{sdaw:eg} through the  process $e^- \gamma \rightarrow W^- Z
\nu_e$, that will be able to  lead to constraints $g^Z_5 \lesssim 0.12$
for a center-of-mass energy of  500 GeV and an integrated luminosity of
$10$ fb$^{-1}$. It is interesting to notice that the bounds obtained in
the above processes originate from the direct tree-level contributions
of the anomalous interaction.

In this work we examine the capability of the next generation of $e^+e^-$
colliders operating in the $\gamma\gamma$ mode to place direct bounds on
the effective operator (\ref{lag:gz5}) through the  reaction $\gamma
\gamma \rightarrow W^+ W^- Z$ \cite{our,our:ano}. In a $\gamma\gamma$
collider, this process exhibits tree-level contributions from the
anomalous interaction (\ref{lag:gz5}) and contains the minimum number of
final state particles. We show that for a center-of-mass of $500$
($1000$) GeV it is possible to obtain bounds $g^Z_5 \lesssim 0.15$ ($4.5
\times 10^{-2}$) for an integrated luminosity of $10$ fb$^{-1}$.

\section{Results}

The most promising mechanism to generate hard photon beams
in an $e^+ e^-$ linear collider is laser backscattering.  Assuming
unpolarized electron and laser beams,  the backscattered photon
distribution function \cite{laser} is
\begin{equation}
F_{\gamma/e}  (x,\xi) \equiv \frac{1}{\sigma_c} \frac{d\sigma_c}{dx} =
\frac{1}{D(\xi)} \left[ 1 - x + \frac{1}{1-x} - \frac{4x}{\xi (1-x)} +
\frac{4
x^2}{\xi^2 (1-x)^2}  \right] \; ,
\label{f:l}
\end{equation}
with
\begin{equation}
D(\xi) = \left(1 - \frac{4}{\xi} - \frac{8}{\xi^2}  \right) \ln (1 + \xi) +
\frac{1}{2} + \frac{8}{\xi} - \frac{1}{2(1 + \xi)^2} \; ,
\end{equation}
where $\sigma_c$ is the Compton cross section, $\xi \simeq 4
E\omega_0/m_e^2$, $m_e$ and $E$ are the electron mass and energy
respectively, and $\omega_0$ is the laser-photon energy. The quantity
$x$ stands for the ratio between the scattered photon and initial
electron energy and its maximum value is
\begin{equation}
x_{\text{max}}= \frac{\xi}{1+\xi} \; .
\end{equation}
In what follows, we assume that the laser frequency is such that $\xi
= 2(1 +\sqrt{2})$, which leads to the hardest possible spectrum of
photons with a large luminosity.

The cross section for $W^+W^-Z$ production via $\gamma\gamma$ fusion can
be  obtained by folding the elementary cross section  for the
subprocesses $\gamma\gamma \rightarrow W^+W^-Z$ with the photon-photon
luminosity ($dL_{\gamma\gamma}/dz$), {\it i.e.},
\begin{equation}
d\sigma (e^+e^-\rightarrow \gamma\gamma \rightarrow WWZ)(s) =
\int_{z_{\text{min}}}^{z_{\text{max}}} dz ~ \frac{dL_{\gamma\gamma}}{dz} ~
d \hat\sigma  (\gamma\gamma \rightarrow WWZ) (\hat s=z^2 s) \; ,
\end{equation}
where $\sqrt{s}$ ($\sqrt{\hat{s}}$) is the $e^+e^-$ ($\gamma\gamma$)
center-of-mass energy, $z^2= \tau \equiv \hat{s}/s$, and the
photon-photon luminosity is
\begin{equation}
\frac{d L_{\gamma\gamma}}{dz} = 2 ~ z  ~
\int_{z^2/x_{\text{max}}}^{x_{\text{max}}} \frac{dx}{x}
F_{\gamma/e} (x,\xi)F_{\gamma/e} (z^2/x,\xi) \; .
\label{lum}
\end{equation}

The analytical calculation of the cross section for the subprocess
 $\gamma\gamma \rightarrow W^+W^-Z$ requires the evaluation of 12
Feynman diagrams in the unitary gauge and it is very lengthy and tedious
 despite being straightforward.  We evaluated numerically the
 helicity amplitudes for this process using the techniques outlined in Refs.\
 \cite{bar:num,zep:num} in order to obtain our results in an efficient
and reliable way. As a check of our results, we explicitly verified that
the amplitudes were Lorentz and $U(1)_{\text em}$ invariant. The phase
space integrations were performed numerically using the Monte Carlo
routine VEGAS \cite{lepage}.

The total cross section for the process $\gamma \gamma \rightarrow
W^+ W^- Z$ is a quadratic function of the anomalous coupling
$g^Z_5$, {\it i.e.}
\begin{equation}
\sigma_{\text{tot}} = \sigma_{\text{sm}} + g^Z_5~ \; \sigma_{\text{int}}
+  (g^Z_5)^2~ \; \sigma_{\text{ano}} \; ,
\label{base}
\end{equation}
where $\sigma_{\text{sm}}$ stands for the SM cross section \cite{our}
and $\sigma_{\text{int}}$ ($\sigma_{\text{ano}}$) is the interference
(pure anomalous) contribution. We evaluated these contributions for
unpolarized backscattered photons  imposing that the polar angles of the
produced vector bosons with the beam pipe are larger than $10^\circ$. In
Table \ref{sigmas:z}, we present our results for several $e^+e^-$
center-of-mass energies. The interference term vanishes since the
anomalous amplitude has a phase of $90^\circ$ with respect to the
standard model amplitude for an unpolarized initial state.

In order to quantify the effect of the new couplings, we defined
the  statistical significance ${\cal S}$ of the anomalous signal
\begin{equation}
{\cal S} = \frac{|\sigma_{\text tot} -
\sigma_{\text sm}|}{\sqrt{\sigma_{\text sm}}} \;
\sqrt{\cal L} \; ,
\label{sig}
\end{equation}
which can be easily evaluated using the parametrization (\ref{base})
with the coefficients given in Table \ref{sigmas:z}.  We list in Table
\ref{const} the values of the anomalous couplings that correspond to a
$3\sigma$ effect in the total cross section for the different
center-of-mass energies of the associated $e^+e^-$ collider, assuming an
integrated luminosity ${\cal L}= 10$ fb$^{-1}$. From this table, we can
learn that a $\gamma\gamma$ collider leads to bounds on $g^Z_5$ that are
better than the ones that can be obtained in the usual $e^+e^-$ mode.
Moreover, $\gamma\gamma$ and $e\gamma$ collider lead to similar
constraints on $g^Z_5$.

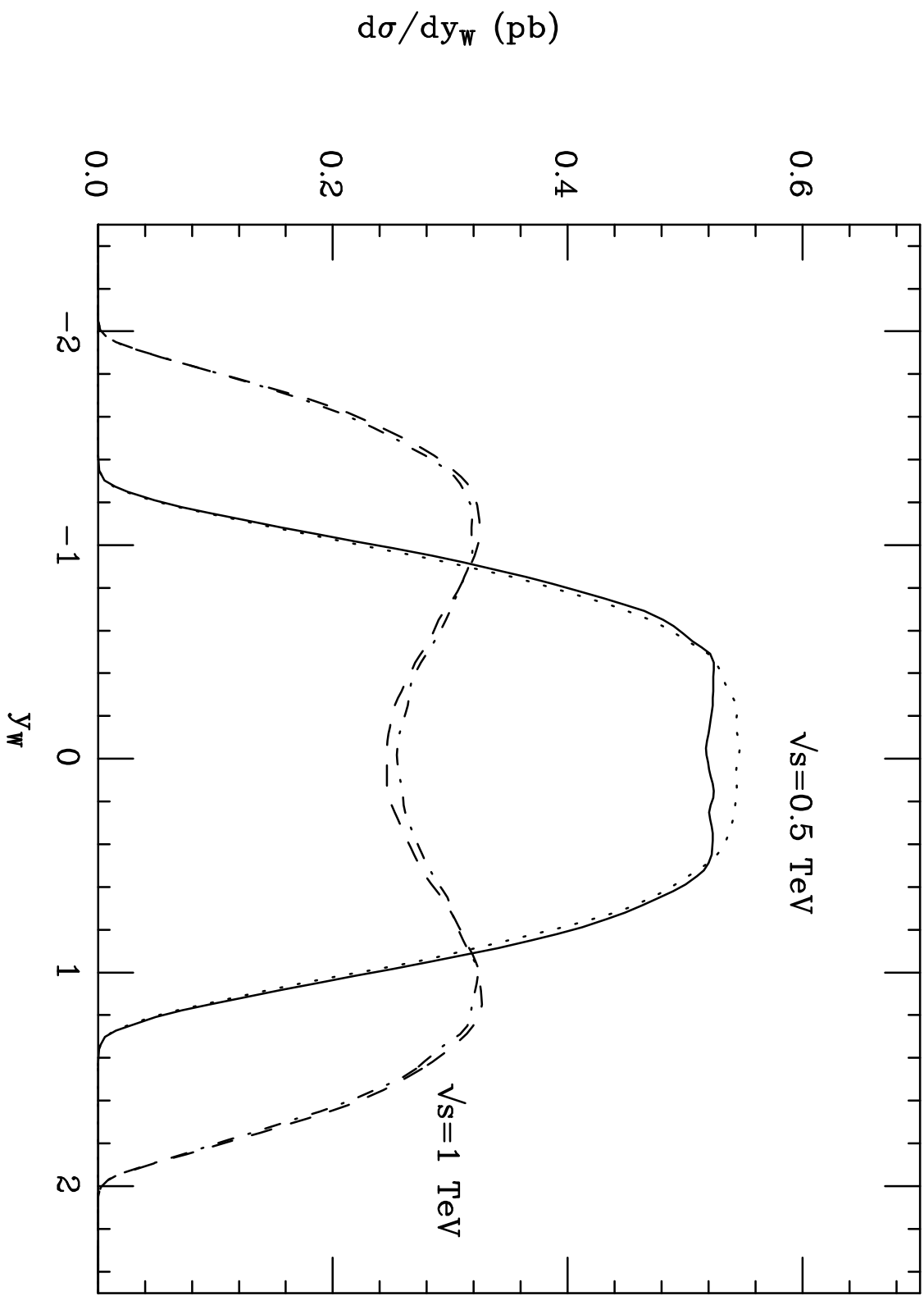
\begin{figure}
\protect
\caption{
  Normalized rapidity distribution of the produced $W^\pm$. The solid
 (dashed) stands for the SM prediction and the dotted
 (dot-dashed) one represents the results for $g^Z_5 = 0.15$ ($4.5
 \times 10^{-2}$) at a center-of-mass energy of $0.5$ ($1$) TeV.
}
\label{fig:1}
\end{figure}

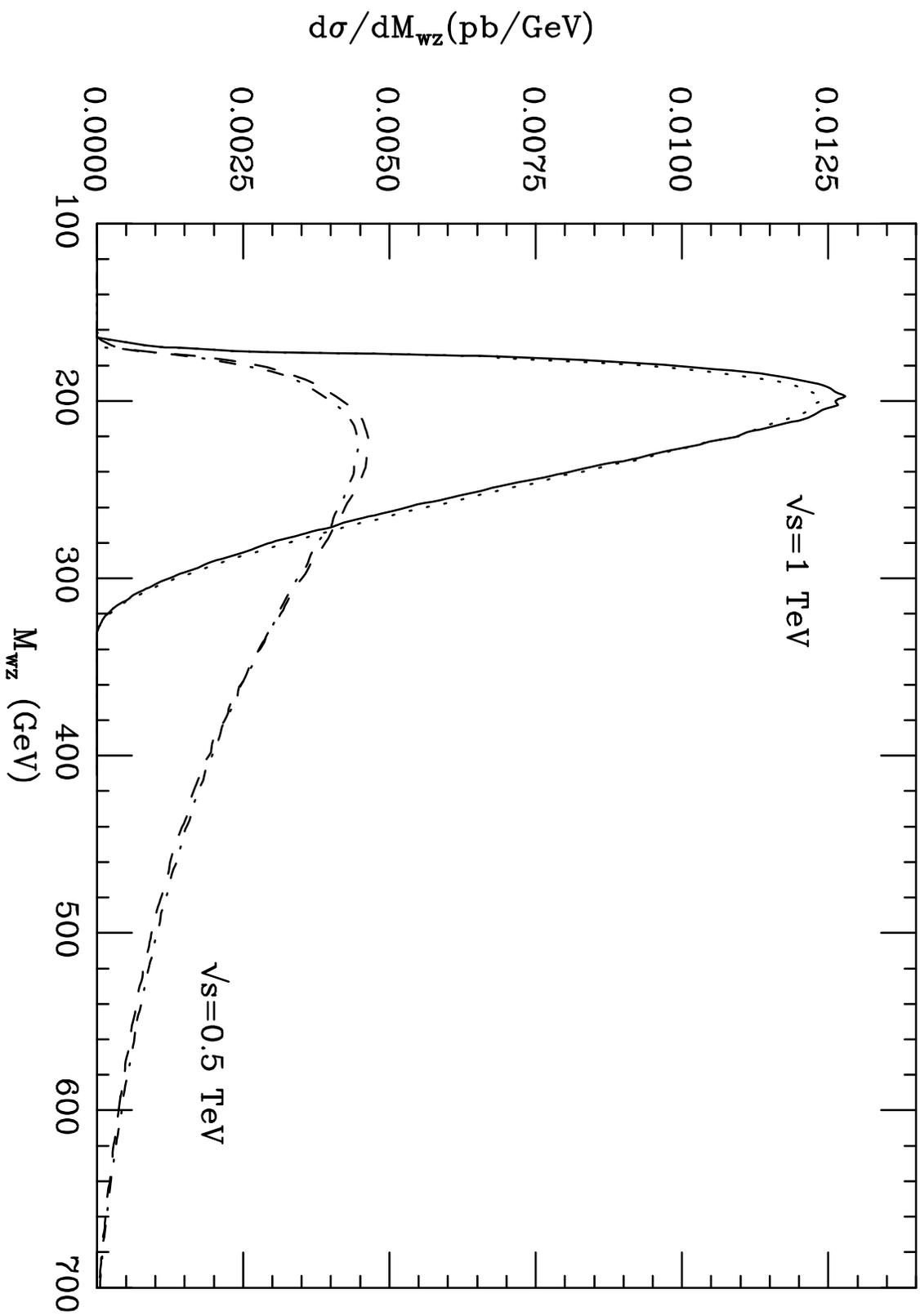
\begin{figure}
\protect
\caption{
  Normalized invariant mass distribution of pairs $W^\pm Z$.
  The conventions are the same as in Fig.\ \protect\ref{fig:1}.
}
\label{fig:2}
\end{figure}

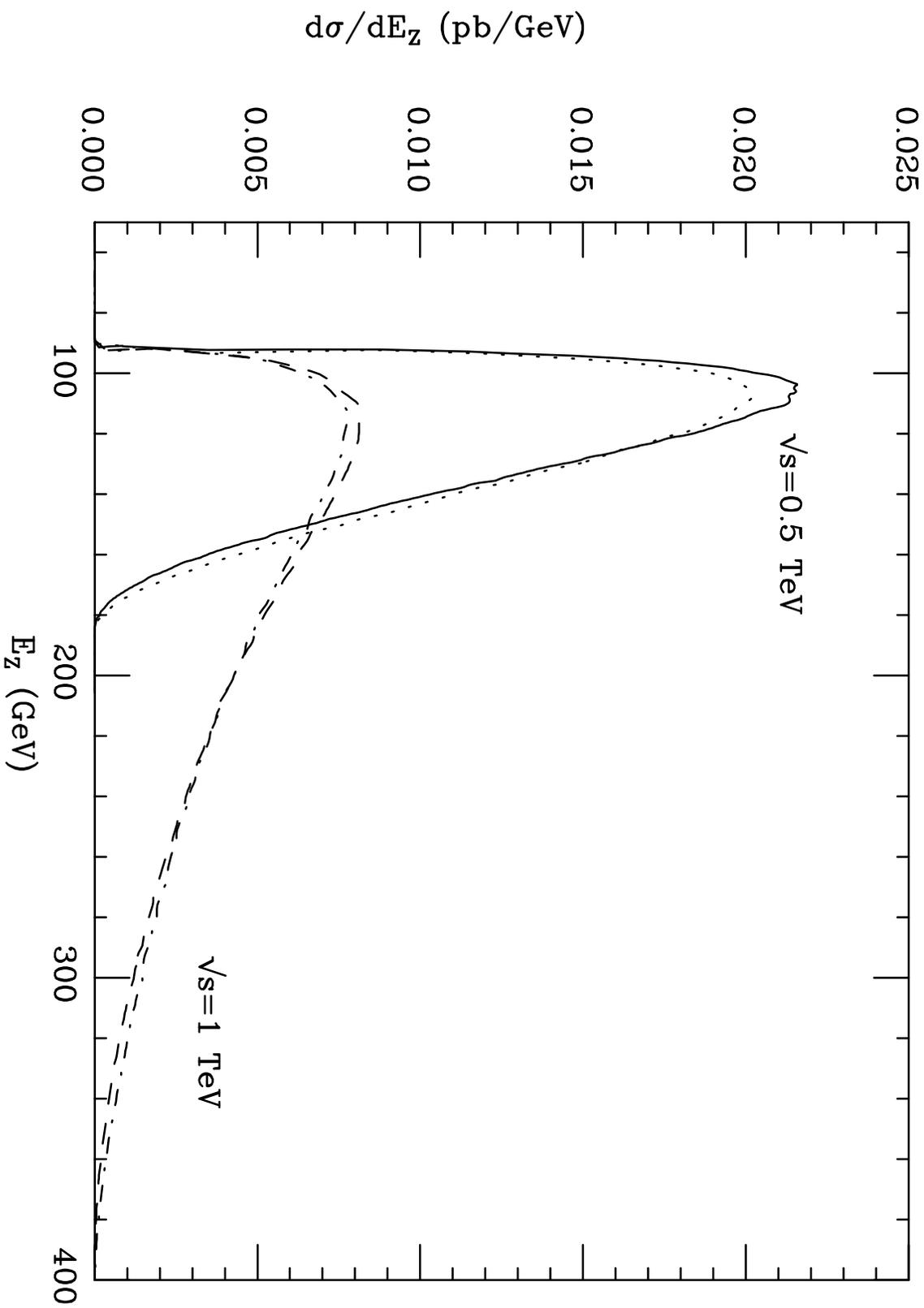
\begin{figure}
\protect
\caption{
  Normalized energy distribution of the $Z$ boson.
  The conventions are the same as in Fig.\ \protect\ref{fig:1}.
}
\label{fig:3}
\end{figure}

The kinematical distributions of the final state particles can be used,
at least in principle, to increase the sensitivity of the $\gamma\gamma$
reactions to anomalous interactions, improving consequently the bounds
on them. In order to reach a better understanding of the effects of the
anomalous interaction (\ref{lag:gz5}) in the reaction $\gamma \gamma
\rightarrow W^+ W^- Z$, we present in Fig.\ \ref{fig:1}--\ref{fig:3}
various representative distributions of the final state gauge bosons,
adopting the values of the anomalous coupling constants that lead to a
$3\sigma$ deviation in the total cross section.

In Fig.\ \ref{fig:1} we show the normalized distribution in the
rapidity $y_W$ of the $W^\pm$ for a center-of-mass energy of $0.5$ and
$1$ TeV.  The distributions for $W^+$ and $W^-$ coincide due to the
absence of the interference term in the cross section. It is
interesting to notice that the anomalous coupling $g^Z_5$ enhances the
production of $W^\pm$ in the central region of the detector, where
they can be more easily reconstructed. Furthermore, increasing the
center of mass energy, the $W$'s tend to populate the high rapidity
region, as it happens in the process $\gamma\gamma \rightarrow W^+
W^-$. Consequently, the cut in the $W$ angle with beam pipe discards a
larger fraction of events at high energies.

The normalized invariant mass distributions of $W^\pm Z$ pairs are
presented in Fig.\ \ref{fig:2} for a center-of-mass energy of $0.5$ and
$1$ TeV. Once again the $W^+Z$ and $W^-Z$ curves coincide.  From this
Figure we can learn that the presence of the anomalous interaction
increases slightly the invariant mass of the $W^\pm Z$ pairs since the
new couplings are proportional to the photon momentum. Moreover,  as the
center-of-mass energy of the collider is increased, the distributions
broaden and  shift toward higher invariant masses.

Figure \ref{fig:3} shows the laboratory energy distribution of the $Z$
gauge boson. As we can see from this figure, the introduction of the
anomalous interaction favors the production of more energetic $Z$
bosons, because of the new momentum-dependent couplings. At lower
center-of-mass energies the distribution is rather peaked around small
values for the energy of the $Z$ boson because of the available
phase space. However, as the center-of-mass energy of the collider
increases, the distributions broaden, exhibiting many $Z$ bosons with
high energies.

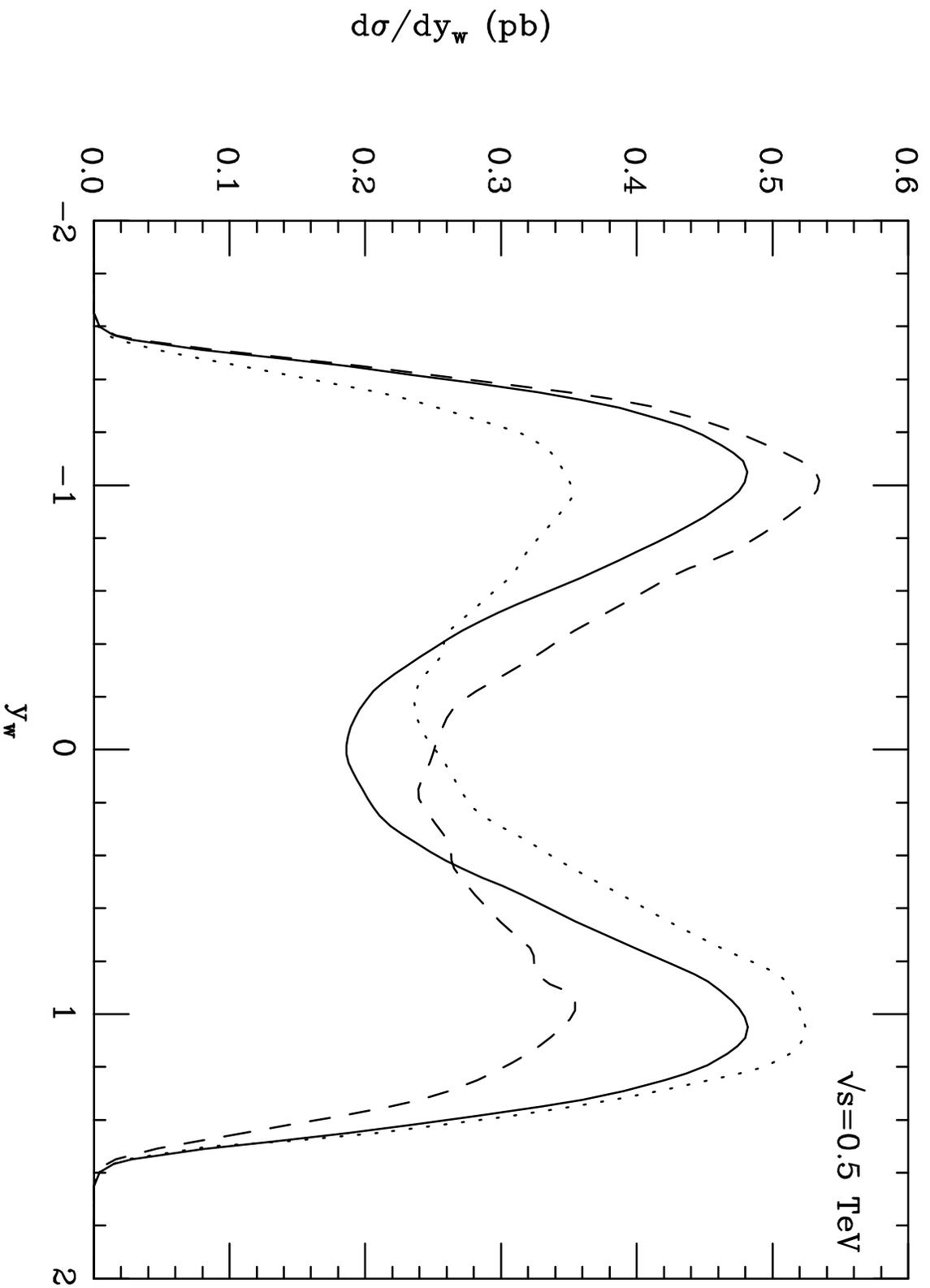
\begin{figure}
\protect
\caption{
  Normalized rapidity distribution of the $W^\pm$ bosons for the
subprocess $\gamma \gamma \rightarrow W^+W^-Z$ at
$\protect\sqrt{\hat{s}} = 0.5$ TeV, using $g_5^Z = 0.15$ and that one
photon is left-handed while the other is right-handed. The solid line
stands for the SM result, while the dotted (dashed) line represents
the anomalous result for the $W^+$ ($W^-$) rapidity.
}
\label{fig:4}
\end{figure}

Up to this point we were able to demonstrate that a $\gamma\gamma$
collider can reveal the existence of an anomalous interaction such as the
one described by (\ref{lag:gz5}). However, the determination that the
anomalous events are because of this interaction is a much harder task. In
principle this could be done through the study of kinematical
distributions. Notwithstanding, several anomalous interactions lead to
distributions similar to the ones that we presented, see for instance
Ref.\ \cite{our:ano}. The effective operator (\ref{lag:gz5}) could be
singled out through the forward-backward asymmetry associated to
parity violation, however, this does not happen for unpolarized
photons since the interference term vanishes, see Table \ref{sigmas:z}.
Therefore, in order to determine which anomalous interaction is
responsible for the anomalous events we must employ polarized
backscattered photons. As an illustration, we show in Fig.\
\ref{fig:4} the normalized $W^\pm$ rapidity distributions for the
subprocess $\gamma \gamma \rightarrow W^+ W^- Z$ with $\sqrt{\hat{s}}
= 0.5$ TeV, assuming that one photon has a left-handed polarization
while the other is right-handed. As we can see from this figure, the
rapidity distribution for the $W^+$ and $W^-$ do not coincide, despite
the result being clearly $CP$ invariant. This a feature unique to the
anomalous interaction (\ref{lag:gz5}).

\section{Conclusions}

We analyzed in this work the capability of an $e^+e^-$ collider
operating in the $\gamma\gamma$ mode to unravel the existence of the
anomalous interaction (\ref{lag:gz5}). We demonstrated that for a
center-of-mass energy of $0.5$ ($1$) TeV and an integrated luminosity
of $10$ fb$^{-1}$, the study of the reaction $\gamma\gamma \rightarrow
W^+ W^- Z$ can lead to bounds $|g^Z_5| \le 0.15$ ($4.5 \times
10^{-2}$).  These bounds are similar to the ones that can be obtained
in the $e\gamma$ mode of the collider and are better than the one
steaming from the usual $e^+e^-$ mode. Moreover, at higher energies
the luminosity of $\gamma\gamma$ colliders can be larger than the
corresponding $e\gamma$ because of problems in the construction this last
mode \cite{berk}.  Consequently, the $\gamma\gamma$ mode will be the
most powerful one to analyze the $g^Z_5$ anomalous coupling.

%**********
\acknowledgments

This work was partially supported by the U.S. Department of Energy
under Contract No. DE-FG02-95ER40896, by the University
of Wisconsin Research Committee with funds granted by the Wisconsin
Alumni Research Foundation, by Conselho Nacional de Desenvolvimento
Cient\'{\i}fico e Tecnol\'ogico (CNPq), and by Funda\c{c}\~ao de
Amparo \`a Pesquisa do Estado de S\~ao Paulo (FAPESP).

%**********
% REFERENCES
%**********

%**********
%Tables
%**********

\begin{table}[htb]
\begin{center}
\begin{tabular}{|c|c|c|c|}
$\protect\sqrt{s}$
&0.5 TeV& 1 TeV& 2.0 TeV
\\ \hline
$\sigma_{\text{sm}}$
& 18.7 & 238.  & 548.
\\ \hline
$\sigma_{\text{int}}$
& 0 & 0 & 0
 \\  \hline
$\sigma_{\text{ano}}$
& 179. & $7.27\times 10^3$ & $142. \times 10^3$
\\
\end{tabular}
\end{center}
\caption{Cross sections
$\sigma_{\text{sm}}$, $\sigma_{\text{int}}$, and
$\sigma_{\text{ano}}$ in fb.}
\label{sigmas:z}
\end{table}

%%%%%%%%

%{\tiny
\begin{table}[htb]
\begin{center}
\begin{tabular}{|c|c|c|c|}
$\protect{\sqrt{s}}$
& $0.5$ TeV & $1$ TeV & $2$ TeV
\\ \hline
$g^Z_5$
& $(-0.15, 015)$ & $(-4.5\times 10^{-2},4.5\times 10^{-2})$
& $(-1.2\times10^{-2},1.2\times10^{-2})$
\\ \hline
$\Delta \sigma$
& 4.14 & 14.6 & 22.2
\\
\end{tabular}
\end{center}
\caption{ Allowed intervals of $g^Z_5$ for an effect smaller
than $3\sigma$ in the total cross section. We also
exhibit the difference ($\Delta\sigma$) between the anomalous
cross sections and the SM ones in fb for a $3\sigma$ effect.}
\label{const}
\end{table}
%}

%**********
% FIGURES
%**********

%%%%%%

%%%%%%

%%%%%%

%%%%%%

\end{document}